\documentclass[showpacs,preprintnumbers,amsmath,amssymb,APSl,prd,nofootinbib,superscriptaddress, onecolumn]{revtex4}
\usepackage[dvips]{graphicx}
\usepackage{amssymb}
\usepackage{amsmath}
\usepackage{graphicx,color}
\usepackage{amsfonts}
\usepackage{bm}
\usepackage{cancel}
\usepackage{comment}
\usepackage{hyperref}
\usepackage{babel}

\begin{document}

\title{Dymnikova GUP-corrected black holes}

\author{G. Alencar \footnote{E-mail: geova@fisica.ufc.br}}\affiliation{Departamento de F\'isica, Universidade Federal do Cear\'a, Caixa Postal 6030, Campus do Pici, 60455-760 Fortaleza, Cear\'a, Brazil}

\author{Milko Estrada\footnote{E-mail: milko.estrada@gmail.com}}\affiliation{Facultad de Ingenier\'ia, Ciencia y Tecnolog\'ia, Universidad Bernardo O'Higgins, Santiago, Chile.}

\author{C. R. Muniz\footnote{E-mail: celio.muniz@uece.br}}\affiliation{Universidade Estadual do Cear\'a (UECE), Faculdade de Educa\c c\~ao, Ci\^encias e Letras de Iguatu, 63500-000, Iguatu, CE, Brazil.}

\author{Gonzalo J. Olmo
	\footnote{E-mail: gonzalo.olmo@uv.es}}
	\affiliation{Departamento de F{\'i}sica Te{\'o}rica and \textit{IFIC}, Centro Mixto Universidad de Valencia - \textit{CSIC}. Universidad de Valencia, Burjassot-46100, Valencia, Spain.}
 \affiliation{Departamento de F\'isica, Universidade Federal do Cear\'a, Caixa Postal 6030, Campus do Pici, 60455-760 Fortaleza, Cear\'a, Brazil}

\begin{abstract}

We consider the impact of Generalized Uncertainty Principle (GUP) effects on the Dymnikova regular black hole. The minimum length scale introduced by the GUP modifies the energy density associated with the gravitational source, referred to as the Dymnikova vacuum, based on its analogy with the gravitational counterpart of the Schwinger effect. We present an approximated analytical solution (together with exact numerical results for comparison) that encompasses a wide range of black hole sizes, spanning from microscopic (Planckian and sub-Planckian) to macroscopic scales, 
whose properties crucially depend on the ratio between the de Sitter core radius and the GUP scale. The emergence of a wormhole inside de Sitter core in the innermost region of the object is one of the most relevant features of this family of solutions. Our findings demonstrate that these solutions remain singularity free, confirming the robustness of the Dymnikova regular black hole under GUP corrections. Regarding energy conditions, we find that the violation of the strong, weak, and null energy conditions which is characteristic of the pure Dymnikova case does not occur at Planckian scales in the GUP corrected solution. This contrast suggests a departure from conventional expectations and highlights the influence of quantum corrections and the GUP in modifying the energy conditions near the Planck scale.

\end{abstract}


\maketitle
\section{Introduction}

Black holes have been a source of fascination for scientists and astronomers, capturing their imaginations for decades. Initially theorized in the realm of General Relativity (GR) over a century ago by Schwarzschild \cite{Schwar}, these enigmatic entities are thought to form from the remnants of massive stars that have undergone gravitational collapse \cite{Oppy}. In addition, they can exist as supermassive objects at the centers of galaxies \cite{Korm}, and also as primordial objects produced by large density fluctuations in the early universe \cite{Ragavendra:2023ret}.

Recent advances in astronomical imaging techniques, such as the Event Horizon Telescope, have enabled us to capture groundbreaking images of astrophysical compact objects, providing unprecedented insights into their properties and behavior. Two remarkable examples of these images are the ones of the center of the M87 galaxy \cite{Kazun} and Sagittarius A*, at the center of our Milky Way galaxy \cite{Kazun1}. Such images have revealed luminosity profiles of accretion disks  that are in excelent agreement with the expectations from black holes, strongly supporting in this way the existence of black holes and offering a visual representation of these objects.

One intriguing area of research focuses on the comparative analysis of modified gravity theories with GR. This analysis goes beyond classical tests applicable to weak fields and relies on the examination of black hole images \cite{Vagnozzi:2022moj} and the detection of gravitational waves resulting from the merging of compact objects \cite{Steer}. If these observations can be further enhanced or refined, they have the potential to provide invaluable insights into the nature of gravity and test the validity of different gravitational theories. These theoretical and observational studies have the potential to significantly advance our understanding of how new physical principles, such as the ones coming from quantum mechanics, can be integrated into theories of gravity and overcome the predictive power of GR, particularly concerning the study of black holes, including their thermodynamics. One compelling approach is based on the Generalized Uncertainty Principle (GUP), which suggests modifications to the classical description of black holes from an extension of the Heisenberg uncertainty principle, which involves a  deformation parameter related to a minimal fundamental length \cite{Elias}. This parameter can be calculated from theoretical and phenomenological approaches \cite{Amati1, Gross, Amati2, Paffuti, Nozari}.

Regular black holes are distinct from the singular ones, and are known to have different properties. The pioneering work by Bardeen \cite{Bardeen} has paved the way for investigating these objects, which may have a finite and non-zero size for their central region, referred to as the core, instead of an infinitely small singularity \cite{Frolov:2016pav,Ali}. Besides nonsingular BHs with de Sitter cores \cite{Dymnikova:1992ux}, we can also have other configurations with a radial bounce. The inner structure of such BHs is very different from the usual GR solutions, including those of Bardeen type. Understanding the properties of regular solutions in both GR and modified theories is an active area of research \cite{Olmo:2015dba,Olmo:2015axa,Menchon:2017qed,Bejarano:2017fgz}. Numerous models, spanning arbitrary dimensions, have been proposed to shed light on their behavior, including the incorporation of exotic matter as a gravitational source, quantum corrections to classical solutions, and hybrid objects that exhibit both singular and regular characteristics, often referred to as ``black bounces'' \cite{Eloy,Stephano,Bazeia:2015uia,Maluf:2018ksj, Maluf:2022jjc, Maluf:2022qfc, Hugo, Simpson:2018tsi, Furtado:2022tnb, Lima:2022pvc}.

In this direction, in the decade of 1990, Dymnikova found a regular black hole \cite{Dymnikova:1992ux} which realizes the Gliner proposal that a de Sitter core would avoid the singularity \cite{Gliner1966, Gliner1970}. According to \cite{Dymnikova:1996plb,Ansoldi:2008jw}, the $d=4$ Dymnikova density profile can be seen as the gravitational analog of the Schwinger effect. In another direction, a GUP correction to the Schwinger effect was obtained in \cite{Haouat:2013yba,Ong:2020tvo}. With this in mind, some of the present authors recently proposed a GUP correction to the Dymnikova density profile and studied wormholes \cite{Estrada:2023pny}. However, the study of black holes with this newly proposed source has not been done yet. 

In this paper, we propose to fill this gap and explore how the introduction of the GUP correction of Ref. \cite{Estrada:2023pny} impacts the properties of the regular Dymnikova black hole. We will see that the de Sitter core structure is preserved under the effects of the GUP correction but the topological aspects of the solution are strongly affected due to the emergence of a wormhole at the innermost region of the object. The size of this wormhole is determined by the GUP correction parameter and one expects it to be of microscopic size, much smaller than the de Sitter core, though it can be tuned to make them comparable in size, which has a nontrivial impact on the 
properties of the resulting solution. 

The paper is organized as follows: Section II reviews the Dymnikova model of regular black holes and introduces the GUP correction. In Section III, we obtain the Dymnikova-corrected black hole solution and investigate some properties of this solution. Finally, in Section IV, we present our conclusions.

\section{Dymnikova's regular black hole and GUP correction}

According to \cite{Dymnikova:1996plb,Ansoldi:2008jw}, the $d=4$ energy density profile associated with the Dymnikova vacuum \cite{Dymnikova:1992ux} can be seen as the gravitational analog of the electron-positron pair production rate, $\Gamma\sim \exp{(-E_c/E)}$, in the vacuum -- the so-called Schwinger effect. This high-energy QED phenomenon is associated with applying an intense uniform electric field ($E$) that results in vacuum polarization and the corresponding production of particle pairs. The critical electric field necessary for abundant pair production is given by $E_c=\pi\hbar m_e^2/e$, where $m_e$ and $e$ are the electron mass and charge modulus, respectively. The gravitational equivalent is heuristically obtained by making the association of the electric field with the gravity tension characterized by a curvature term, namely 
\begin{equation}\label{identification}
 E\sim r^{-3},\; \frac{E_c}{E} =\frac{r^3}{r_g r_0^2} \ ,   
\end{equation}
where, $r_g=2M$ and $r_0$ is related to the curvature of the de Sitter core. With this, we obtain the $d=4$ ``Dymnikova-Schwinger'' density profile given by 
\begin{equation}\label{puredymnikova}
\rho(r)=\rho_0 \exp{\left(-\frac{r^3}{a^3}\right)}.
\end{equation}
In the above expression, we have defined $a^3=r_gr_0^2$. To get the de Sitter core we also must  have $r_0^2=3/(8\pi \rho_0)$. 

The correction to the Schwinger effect associated with the existence of a minimal length $\ell$ was obtained in \cite{Haouat:2013yba,Ong:2020tvo} through the Generalized Uncertainty Principle (GUP). For small $\alpha=\ell^2$, the authors show that 
\begin{equation} \label{smallalpha}
\Gamma\sim \exp{\left(-\frac{A}{E}+B(\alpha) E\right)},
\end{equation}
where $A$, $B(\alpha)$ are constants depending on the mass and charge of the electron, $E$ is the electric field, and $\alpha$ comes from GUP via
\begin{equation}
\Delta x\Delta p\sim \frac{\hbar}{2}\left[1+\frac{\alpha(\Delta p)^2}{\hbar}\right].
\end{equation}
To study black holes, we need the full expression for particle production with GUP correction. The authors of Ref. \cite{Haouat:2013yba} provide this, and it is given by
\begin{equation}\label{fullproduction}
\Gamma\sim \frac{\sinh^2\left(\frac{\pi}{eE\alpha}\sqrt{1-\alpha m^2}\right)}{\cosh^2\left(\frac{\pi}{eE\alpha}\sqrt{1-(\frac{eE\alpha}{2})^2}\right)},
\end{equation}
With the above expression, we can perform a complete system study. To make contact with the analyais of Ref. \cite{Estrada:2023pny}, we will carefully consider the limit of small $\alpha$. In the limit of  $\alpha m^2\ll 1,eE \alpha\ll 1 $ Eq. (\ref{fullproduction}) gives 
\begin{equation}\label{correctaappdymnikova}
\Gamma\sim \exp{\left[-\frac{\pi m^2}{eE}+\frac{\pi\alpha}{4} Ee\right]}.
\end{equation}
To consistently obtain a GUP-Dymnikova density, we must take a step further in the identifications (\ref{identification}). First, the curvature tension that provides the $~1/r^3$ dependence of the Dymnikova model is the Kretschmann scalar $K$. To be more precise, for the Schwarzschild case, we have $\sqrt{K}\sim  r_g/r^3$. Therefore, we get the correct dependence in $r_g$. This dependence makes sense since the tension of spacetime must grow with the mass of the source. Next, we can see that we must identify $m^2\sim 1/r_0^2$. With this, we get precisely 
\begin{equation}\label{correctiden}
    \frac{eE}{\pi}\sim \sqrt{K} \sim  r_g/r^3, \;  m^2\sim 1/r_0^2\to \frac{E_c}{E} =\frac{r^3}{r_g r_0^2}. 
\end{equation}
This is exactly the identification \ref{identification} proposed by Dymnikova. The above analysis is crucial since new parameters are entering the theory. We also remember that we must maintain the above identifications to recover the Dymnikova density in the limit $\alpha=0$. By replacing (\ref{correctiden}) in (\ref{correctaappdymnikova}), we obtain that the GUP-Dymnikova density profile is given by
\begin{equation}\label{correctappgup}
\rho(r)= \rho_0 \exp{\left[-\frac{r^3}{a^3}+\alpha\frac{b}{r^3} \right]};\; b=\frac{\pi^2 r_g}{4} 
\end{equation}
Despite having the correct $r$ dependence, the constant in the second term is quite different and can change some of the conclusions of Ref. \cite{Estrada:2023pny}. However, the objective of the present manuscript is to study black holes.

\begin{figure}[!h]
 \centering
    \includegraphics[width=0.6\textwidth]{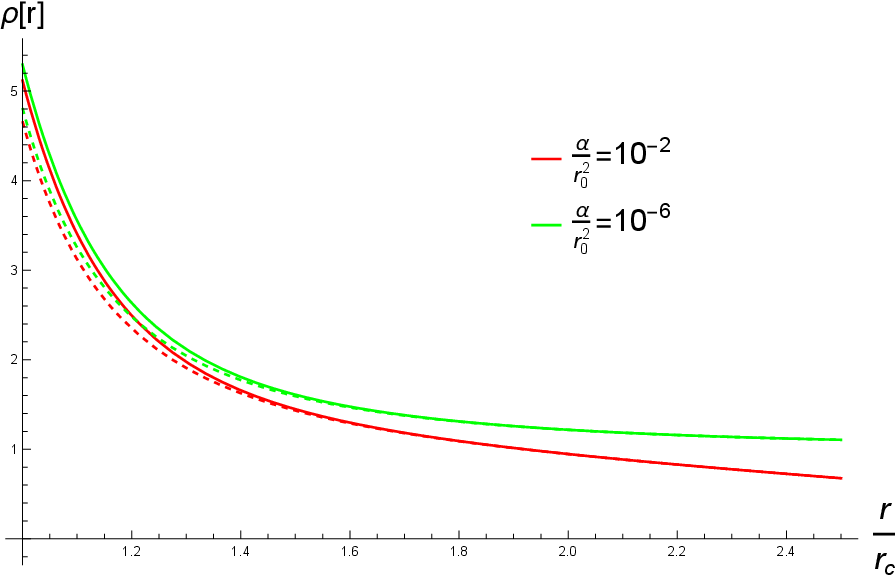}
        \caption{Comparison of the density function (\ref{fulldensity}) (solid curve) with the approximated expression (\ref{correctappgup}) (dashed curve in the same color). The difference is only apparent around $r=r_c$ for any reasonable value of the ratio $\alpha/r_0^2$.}
    \label{DensityPlot}
\end{figure}

As {one can see}, the density profile  (\ref{correctappgup}) is divergent {in the limit} $r\to0$, {which} would lead to a singularity at the origin. {As a result}, {one may think that}  {the introduction of} a minimal length {may} break the regularity of the Dymnikova black hole. {This, however,  is highly counterintuitive, because one would expect that a refinement of the quantum properties due to the GUP should not spoil the robustness of the de Sitter core, which supposedly represents the bulk of the quantum gravitational regularization}. {In this sense,  one should bear in mind that} the above expression is just an approximation valid  only for large $r$, {and that the full expression (\ref{fullproduction}) should be considered at short distances}.  By using our identification (\ref{correctiden}) the density {that follows from (\ref{fullproduction}) is}
\begin{equation}\label{fulldensity}
\rho(r)= \rho_0\frac{\sinh^2\left(\frac{ r^3}{r_g\alpha}\sqrt{1-\alpha \frac{1}{ r_0^2}}\right)}{\cosh^2\left(\frac{ r^3}{r_g\alpha}\sqrt{1-\alpha^2(\frac{\pi r_g}{2r^3})^2}\right)}=\rho_0\frac{\sinh^2\left(\frac{ \pi r^3}{2 r_c^3}\sqrt{1-\alpha \frac{1}{ r_0^2}}\right)}{\cosh^2\left(\frac{ \pi r^3}{2 r_c^3}\sqrt{1-(\frac{r_c}{r})^6}\right)}.
\end{equation}
In the last equality, we have defined $r_c^3=\pi r_g\alpha/2 $ ({recall that $\alpha\equiv \ell^2$}). In the limit $r_c\to0$, (\ref{fulldensity}) recovers the Dymnikova density \ref{puredymnikova}.  For small $r_c$, Eq. (\ref{fulldensity}) also recovers Eq. (\ref{correctappgup}), {since it behaves as }
\begin{equation}\label{eq:rhoapprox}
 \rho(r)\approx \tilde{\rho}_0 e^{(\lambda-1)\pi x^3+\frac{\pi}{2x^3}}    \ ,
\end{equation}
{where $x\equiv r/r_c$, $\tilde{\rho}_0\equiv  \rho_0 \sinh^2\left(\frac{\lambda \pi}{2}\right)$, and $\lambda=\sqrt{1-\alpha/r_0^2}$.} {In fact, a careful comparison of the exact and approximate expressions shows that they are very similar everywhere in the interval $r\ge r_c$, which is where the exact formula (\ref{fulldensity}) is defined (see Fig.\ref{DensityPlot}). Thus, the most important aspect of Eq. (\ref{fulldensity}) is that it restricts the domain of definition of the radial variable to the region $r\ge r_c$, avoiding in that way the disturbing limit $r\to 0$, which should be regarded as nonphysical. Note also that to have a decaying exponential, the factor $\lambda-1$ in Eq.(\ref{eq:rhoapprox}) must be negative, which requires $\alpha>0$.}

In order to understand the meaning of the restriction $r\ge r_c$ (or $x>1$), it is useful to have a look first at Eq.(\ref{fullproduction}). The inclusion of a minimal length in the electromagnetic scenario implies the existence of a maximum electric field intensity. This is reminiscent of the Born-Infeld nonlinear theory of electrodynamics, where a square root structure in the action acts as a mechanism to bound from above the electric field and regularize the self-energy of point-particles. When this is translated to the gravitational sector, the square root in the denominator of (\ref{fulldensity}) implies a bound on the maximum curvature, which is achieved by limiting the minimal area of the $2-$spheres. This is why the obtained solution must satisfy $r\ge r_c$. Restrictions of this type have already been observed in the Born-Infeld theory of gravity \cite{BeltranJimenez:2017doy}, where one finds nonsingular bouncing cosmologies and regular black holes. The regularity of those black holes occurs by the same mechanism found here, namely, by the emergence of a minimal $2-$sphere that represents the throat of a wormhole. 

Above, we have considered the system's behavior for small $\alpha$. However, there is another interesting case to be studied. Note that the square root in the numerator of (\ref{fulldensity}) must also be real. This implies that the range of $\alpha$ is limited to $0<\alpha<r_0^2$. Therefore, we also have an upper value of $\alpha$. With this, we also get that the range of the critical radius is limited to $0<r_c^3<\pi r_g r_0^2/2$. Below, we summarize the ranges of our parameters
\begin{equation}\label{ranges}
0<\alpha<r_0^2; \;  0<r_c^3< \frac{\pi r_g r_0^2}{2}.
\end{equation}
First we note the curious fact that the minimal GUP length can never be bigger than the de Sitter core. Second, close to the upper limit $\alpha=r_0^2$ ($\lambda\approx 0$), our density simplifies to
\begin{equation}\label{fulldensityupper}
\rho(r)=\rho_0\left(\frac{ \pi r^3}{2 r_c^3}\right)^2\lambda^2\text{sech}^2\left(\frac{ \pi r^3}{2 r_c^3}\sqrt{1-\left(\frac{r_c}{r}\right)^6}\right).
\end{equation}

In the next section, we analyze some consequences of  (\ref{fulldensity}).

\section{The solution and its properties}

Regular black hole solutions with de Sitter cores are generally sourced by nonlinear theories of electrodynamics (NEDs), for which the stress-energy tensor satisfies $p_r=-\rho$  and $p_\theta=K(\rho)$, where $\rho=-(\varphi - 2X \varphi_X)/8\pi$, $p_\theta=\varphi(X)/8\pi$, and $\varphi(X)$ is the NED Lagrangian, which is a function of the electromagnetic invariant $X\equiv -F_{\mu\nu}F^{\mu\nu}/2$. For this type of sources, the line element can be written as
\begin{equation}
    ds^2=-f(r)dt^2+\frac{dr^2}{f(r)}+r^2d\Omega^2,
\end{equation}
{and the Einstein equations lead to }
\begin{equation}\label{eq:f_edo}
  (r(1-f))'=\frac{\kappa^2}{4\pi} r^2 \rho  
\end{equation}
where $\kappa^2=8\pi G$ and a prime indicates derivation with respect to the radial coordinate. For simplicity, we will set Newton's constant to unity. Though it is not possible to obtain an analytic solution to the Einstein equations using the energy density \eqref{fulldensity}, {an excellent approximation can be found using the expression (\ref{eq:rhoapprox})}. 
{The solution to Eq.(\ref{eq:f_edo}) can thus be approximated as}
\begin{equation}
    f(r)=1-\frac{2\tilde{\rho}_0r_c^2}{x}m(x) \ ,
\end{equation}
where $m(x)$ is given by 
\begin{equation}
    m(x)\equiv \int_1^x dx x^2 e^{(\lambda-1)\pi x^3+\frac{\pi}{2x^3}}=\frac{1}{3}\int_1^{x^3} dz e^{(\lambda-1)\pi z+\frac{\pi}{2z}} \ .
\end{equation}
It is worth noting that the function $m(x)$ only depends on the parameter $\lambda\equiv \sqrt{1-\alpha/r_0^2}$, which represents the relation between the GUP scale $\alpha$ and the de Sitter core of the original solution. Additionally, the factor $2\tilde{\rho}_0r_c^2$ can be expressed as $3 \pi ^{2/3} \left(1-\lambda ^2\right) \sqrt[3]{\frac{M^2}{\alpha }}$, which represents the relation between the GUP scale and the Schwarzchild radius. Thus, the solution is characterized by just two free parameters, namely, $\alpha/r_0^2$ and $M^2/\alpha$.   

As one can see, the factor $e^{(\lambda-1)\pi z}$ in the integrand determines the asymptotic decay of the density profile. Assuming that $(\lambda-1)\approx -\alpha/2r_0^2$ is a very small number, the contribution of this term is negligible for small values of $z$, and one can safely take $e^{(\lambda-1)\pi z}\approx 1$ for not too large values of $z$. Thus, the main contribution from the innermost part of the core comes from the $e^{\frac{\pi}{2z}}$ piece, which rapidly tends to unity away from $z=\pi/2$. With these considerations and a bit of numerical analysis, we find that $m(x)$ can be well approximated by the following expression:
\begin{equation}\label{eq:mass_profile}
    m(x)\approx \frac{1}{3} \left( m_0(\lambda)+\frac{e^{(\lambda-1)\pi x^3}}{\pi(\lambda-1)}+b(\lambda)\text{Ei}\left[(\lambda-1)\pi x^3\right]\right) \ ,
\end{equation}
where 
\begin{eqnarray}
    m_0(\lambda)&\equiv& -\frac{1}{\pi(\lambda-1)}-2.2-1.6 \ln[1-\lambda]  \nonumber \\
    b(\lambda)&\equiv & 1.6+2.2\sqrt{1-\lambda} \ .
\end{eqnarray}

\begin{figure}[!h]
 \centering
    \includegraphics[width=0.6\textwidth]{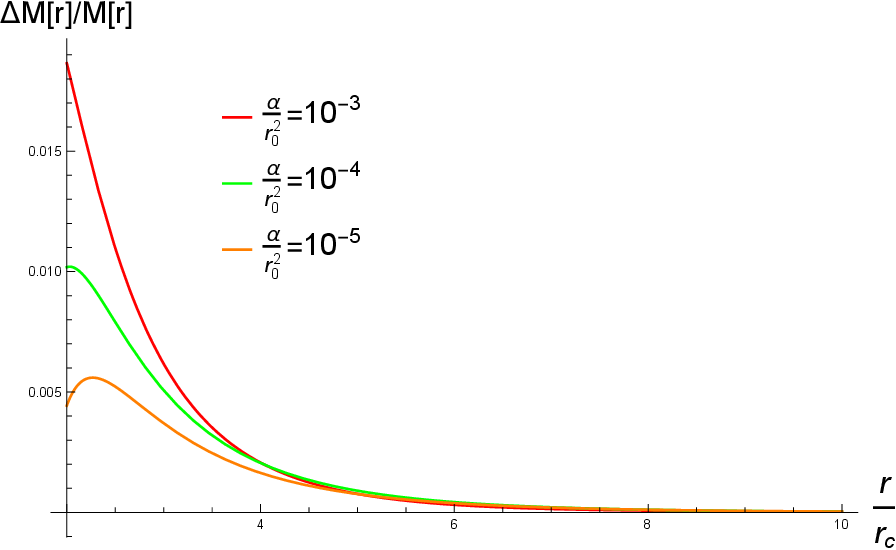}
        \caption{{Relative deviation of the approximated mass function (\ref{eq:mass_profile}) compared to the numerical integration of the density function (\ref{eq:rhoapprox}). Though at $x=1$ the error is of order unity, from $x\ge 2$ onwards the difference is always smaller than $1$ per cent. Since the approximated density function (\ref{eq:rhoapprox}) only departs from the exact one within $x\leq 2$, we can confirm that the mass function (\ref{eq:mass_profile}) is an excellent approximation everywhere except when $x\leq 2$}}
    \label{RelativeError}
\end{figure}

\subsection{Approximation $\alpha\ll r_0^2$}

From the approximated expressions derived above, the metric function $f(r)$ in the limit of small $\alpha/r_0^2$ can be written as: 
\begin{equation} \label{funcionr>>l}
    f(r)=1-\frac{\pi \tilde{\rho}_0 r_g \alpha}{r}\left(-\frac{2.2+1.6\ln\left(\frac{\alpha}{r_0^2}\right)}{3}+\frac{2r_0^2}{3\pi\alpha}\left(1-e^{-r^3/a^3}\right)+\frac{1}{3}\left(1.6+2.2 \frac{\alpha}{2r_0^2}\right)Ei\left[-\frac{r^3}{a^3}\right]\right) \ .
\end{equation}
From this, it is inmediate to compute the limit $\alpha\to 0$, which yields a finite result:
\begin{equation} \label{limit_1}
    f(r)=1-\frac{2 \tilde{\rho}_0 a^3}{3r}\left(1-e^{-r^3/a^3}\right) \ .
\end{equation}
This limit has two characteristic regimes, namely, when $r/a\ll 1$ one obtains a de Sitter core of the form $f(r)\approx 1-{2 \tilde{\rho}_0 r^2}/{3}$, while when $r/a\gg 1$ one recovers the usual Schwarzschild solution $f(r)\approx 1-{2 \tilde{\rho}_0 a^3}/{3r}$, where one should identify the Schwarzschild mass as $M\equiv \tilde{\rho}_0 a^3/3$. As expected, this coincides with the Dymnikova solution. Using the notation in terms of the mass $M$ and splitting the Dymnikova part from the GUP correction, we can finally write the function $f(r)$ in the form
\begin{equation} \label{funcionr_final}
    f(r)=1-\frac{2 M}{r}\left(1-e^{-r^3/a^3}\right)+\left(\frac{\alpha r_g \pi}{a^3}\right)\frac{M}{r}\left(2.2+1.6\ln\left(\frac{\alpha}{r_0^2}\right)-\left(1.6+2.2 \frac{\alpha}{2r_0^2}\right)Ei\left[-\frac{r^3}{a^3}\right]\right) \ .
\end{equation}
This makes it clear that for any length scale $r$ larger than $\alpha^{1/2}$, the geometry is very well approximated by the Dymnikova solution. Note that the coefficient in front of the GUP correction can also be written as $\left(\frac{\alpha r_g \pi}{a^3}\right)=\left(\frac{2r_c^3}{a^3}\right)=\left(\frac{\alpha \pi}{r_0^2}\right)=\pi(1-\lambda^2)$.

\begin{figure}[!h]
 \centering
    \includegraphics[width=0.8\textwidth]{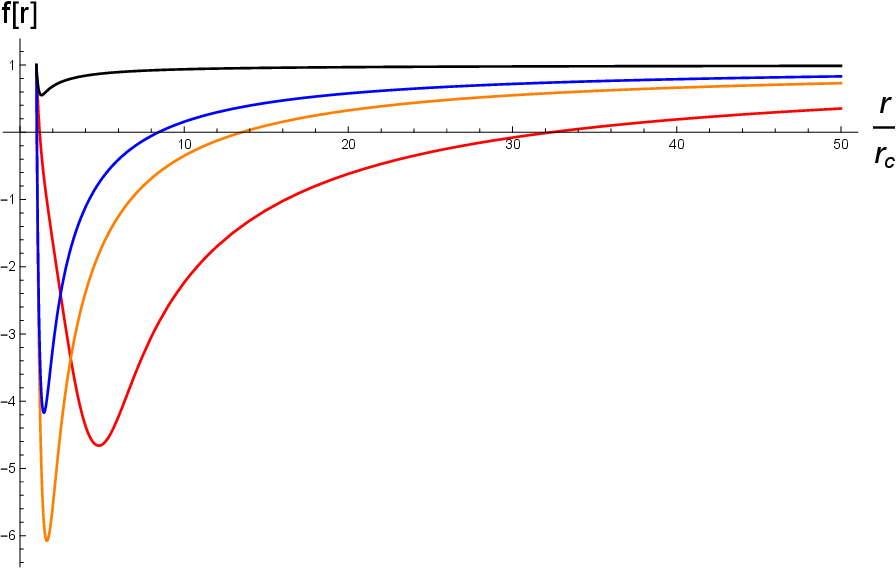}
        \caption{Metric coefficient $f(r)$ given by exact numerical integration. Red curve: $r_0=10\sqrt{\alpha}$, $M=10 r_0$; Orange curve: $r_0=2\sqrt{\alpha}$, $M=10 r_0$; Blue curve: $r_0=1.5\sqrt{\alpha}$, $M=10 r_0$; Black curve: $r_0=1.1\sqrt{\alpha}, M=2r_0$. Only the black curve represents a horizonless configuration (traversable wormhole).}
    \label{Effedeerre1}
\end{figure}

In Fig \ref{Effedeerre1}, we depict the metric coefficient $f(r)$ given by numerical integration of the exact energy density (which is virtually identical to the result provided by Eq. (\ref{funcionr_final})) in order to analyze the presence of horizons. As we can see, microscopic configurations may be horizonless, representing traversable wormholes.

\subsection{Approximation $r\approx r_c$}

In order to investigate the behavior of the solution around the wormhole throat, where $r=r_c$, we expand the full density in Eq. (\ref{fulldensity}) around $r = r_c$ and find an expression of the form $\rho(r)/\rho_0\approx c_1-c_2 (r-r_c)/r_c$, and upon integration we obtain
\begin{equation}\label{r=r_c}
   f(r)\approx 1-\frac{2\rho_0(c_1+c_2)}{3} r^2+\frac{\rho_0 c_2}{2r_c} r^3,
\end{equation}
where 
\begin{eqnarray}\label{c's}
c_1&=& \sinh\left(\frac{\pi \lambda}{2}\right)^2 \ ,\nonumber\\ 
c_2&=& \frac{3\pi}{2}\left(\pi \sinh\left(\frac{\pi \lambda}{2}\right)^2-\lambda \sinh\left({\pi \lambda}\right)\right)  \ .
\end{eqnarray}
As can be seen, this functional form is in excellent agreement with our analysis above of the limit $\alpha\to 0$ when $r\ll a$ using the approximated energy density (\ref{eq:rhoapprox}). The leading order correction of our approximated expansion only misses the contribution due to the constant $c_2$ (recall that $\tilde{\rho}_0=\rho_0 c_1$). Our approximated expressions can only capture the $r^2$ term plus order $r^5$ corrections, missing the $r^3$  correction of the exact solution. In any case, as already shown in Fig.\ref{RelativeError}, beyond $r\approx 2 r_c$ the approximated expressions are valid within a $1\%$ accuracy at least. This analysis of the region $r\approx r_c$ further supports the qualitative and quantitative validity of our analytic approximations.

\subsection{Curvature}

\begin{figure}[!h]
\centering
 \includegraphics[width=0.6\textwidth]{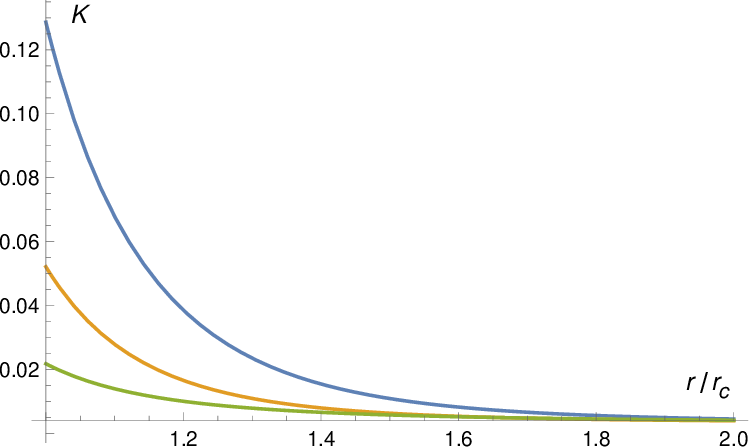}
   \caption{Kretschmann scalar as a function of the ratio between the radial coordinate and the curvature radius, in the linear approximation for $\alpha$, considering $M=20$, $a=15$, and $\alpha=0.015$, in Planck coordinates.}
 \label{Kretsch}
\end{figure}

Now, let's analyze the behavior of the curvature, specifically the Kretschmann scalar denoted by $K(r)$. It is defined as $K(r) = (f'')^2 + \frac{4}{r^2}(f')^2 + \frac{4}{r^4}(1-f)^2$. Even considering the linear approximation for the $\alpha$ parameter in Eq. (\ref{funcionr_final}), the expression for that such a scalar is very involved. Thus, we depict it in the Fig \ref{Kretsch}.
We can notice that when the radial coordinate $r$ tends to infinity, the Kretschmann scalar $K(r)$ approaches zero, indicating the asymptotic flatness of the Dymnikova-GUP metric, as anticipated. It is important to highlight that the Dymnikova-GUP metric remains devoid of singularities for $r\geq r_c$, as long as the parameter $\alpha^{1/2}$ is sufficiently small. This condition guarantees that quantum gravity corrections remain confined to smaller scales, preventing the formation of singularities within the region in which the spacetime can be defined yet. 

From eqs. (\ref{r=r_c}) and (\ref{c's}), in the neighbourhood of $r=r_c$, the Kretschmann scalar is 
\begin{equation}
   K(r)\approx \frac{\rho_0^2 \left[32 c_1^2 r_c^2+16 c_1 c_2 r_c (4 r_c-5 r)+c_2^2 \left(57 r^2-80 r_c r+32 r_c^2\right)\right]}{3 r_c^2}.
\end{equation}
The smoothness of the Kretschmann curvature in the regime where $r\approx r_c$ and $\ell\to 0$ is readily demonstrated, unequivocally also establishing the regularity of the Dymnikova-GUP solution in that region.

We explore next the energy conditions and their potential violations. 

\subsection{Energy conditions}

We now analyze the energy conditions of our solution. In the Dymnikova case, the strong energy condition is always violated. This is expected since it is a regular solution. Here we will analyze modifications introduced by the GUP correction. 

The tangential pressure components, $p_t=p_\theta=p_\phi$ can be computed by using the conservation equation or the Einstein Equation and is given by
\begin{equation}\label{lateral}
  p_t=-\rho-\frac{r}{2} \rho'.
\end{equation}
 
Therefore  $\rho'$ is needed to study the energy conditions. This can be obtained from (\ref{fulldensity}), which yields
\begin{eqnarray}\label{rhoprime}
\rho'(r)&=& \rho_0\frac{3\sinh\left(\frac{ \pi r^3}{2 r_c^3}\lambda\right)\cosh\left(\frac{ \pi r^3}{2 r_c^3}\lambda\right)}{\cosh^2\left(\frac{ \pi r^3}{2 r_c^3}\sqrt{1-(\frac{r_c}{r})^6}\right)}\frac{ \pi r^2}{ r_c^3}\lambda\nonumber\\
&&-\rho_0\frac{3r^5\pi \sinh^2\left(\frac{ \pi r^3}{ r_c^3}\lambda\right)\sinh\left(\frac{ \pi r^3}{2 r_c^3}\sqrt{1-(\frac{r_c}{r})^6}\right)}{ r_c^3\sqrt{r^6-r_c^6}\cosh^3\left(\frac{ \pi r^3}{2 r_c^3}\sqrt{1-(\frac{r_c}{r})^6}\right)}.
\end{eqnarray}
This expression is crucial to study the energy conditions. 

\subsubsection{Weak and Null Conditions}
First, we have that  $\rho+p_r=0$ and $\rho\geq 0$  are automatically satisfied. Therefore, for the null and weak conditions to be satisfied, we need the condition 
\begin{equation} \label{nulllateral}
\rho+p_t\ge 0\to E_w(r)\equiv-\frac{r}{2} \rho'\ge 0,
\end{equation}
where we have used (\ref{lateral}). Employing (\ref{rhoprime}), the above condition takes the form
\begin{eqnarray}\label{weakandnull}
E_w(r)&= &- 3\rho_0 \frac{\sinh\left(\frac{  r^3}{2 r_c^3}\lambda\right)\cosh\left(\frac{ \pi r^3}{2 r_c^3}\lambda\right)}{\cosh^2\left(\frac{ \pi r^3}{2 r_c^3}\sqrt{1-(\frac{r_c}{r})^6}\right)}\frac{ \pi r^3}{ r_c^3}\lambda\nonumber\\
&&+3\rho_0\frac{r^6\pi \sinh^2\left(\frac{ r^3}{ r_c^3}\lambda\right)\sinh\left(\frac{ \pi r^3}{2 r_c^3}\sqrt{1-(\frac{r_c}{r})^6}\right)}{ r_c^3\sqrt{r^6-r_c^6}\cosh^3\left(\frac{ \pi r^3}{2 r_c^3}\sqrt{1-(\frac{r_c}{r})^6}\right)}\ge 0.
\end{eqnarray}
 In the regime $r\gg r_c$ (small $\alpha$), (\ref{weakandnull}) reduces to
$$\frac{3}{2} \left(\frac{r^3}{a^3}+\frac{\alpha a}{r^3}\right)\rho\geq 0.$$
Therefore, the null condition is satisfied, and since we also have $\rho\geq 0$, the weak energy condition is also satisfied.

Let us analyze the case close to $r_c$. For this, we expand the above expression in $\xi=r-r_c$ to get
\begin{equation} \label{linearweak}
   E_w(r)=c_1+c_2\xi 
\end{equation}
with 
\begin{equation}\label{aweak}
c_1=   \frac{3\rho_0}{4} \left(\pi ^2 \sinh ^2\left(\frac{\pi  \lambda }{2}\right)-\pi  \lambda  \sinh (\pi  \lambda )\right).
\end{equation}
and 
\begin{align}\label{bweak}
&c_2=\frac{3\pi^2\rho_0}{4r_c} \left(\pi ^2-3\right)-\frac{3\pi^2\rho_0}{4r_c}  \left(3 \lambda ^2+\pi ^2-3\right) \cosh (\pi  \lambda )\nonumber \\
&+\frac{9\pi\rho_0}{4r_c} \left(\pi ^2-1\right) \lambda  \sinh (\pi  \lambda )
\end{align}
From (\ref{aweak}) and remembering that $0\leq\lambda\leq 1$, we always have that $c_1>0$. Therefore, for $\xi=0$ ($r=r_c$),  the weak and null energy conditions are satisfied. 

Let us now analyze if there is some region in which the conditions are violated. Since (\ref{linearweak}) is linear and $c_1>0$, the conditions are violated for $\xi>-c_1/c_2$, if $c_2<0$. We must be careful when analyzing $c_2$ as a function of $\lambda$. We should remember that both $\lambda$ and $r_c$, depend on $\alpha$. Restoring this in expression \ref{bweak}, close to $\lambda=1$ we have 
$$
c_2\propto -\frac{1}{\alpha^{1/3}} \to \lim_{\alpha \to 0}c_2=-\infty
$$ 
Therefore, in  the limit $\alpha \to 0$, the conditions are violated for all $r$. This is precisely the pure Dymnikova limit.  However, when $\lambda=0$ (and therefore the upper value of $\alpha$), we get $c_2\to 0 $, and the conditions are never violated in the regimen $r_c=r_0$. Therefore the limit $\lambda=0$ is very important, and we must analyze if this conclusion is valid only close to $r=r_c$. For this, we must use (\ref{fulldensityupper}) in (\ref{nulllateral}) to get
\begin{equation}
  E_w(r)=  \lambda^2 \rho_0\frac{3 \pi ^2 r^6 \text{sech}^2\left(\frac{\pi  r^3 }{2 r_c^3}\sqrt{1-\frac{r_c^6}{r^6}}\right)}{8 r_c^9 \sqrt{1-\frac{r_c^6}{r^6}}}\left(\pi  r^3 \tanh \left(\frac{\pi  r^3 }{2 r_c^3}\sqrt{1-\frac{r_c^6}{r^6}}\right)-2 r_c^3 \sqrt{1-\frac{r_c^6}{r^6}}\right)
\end{equation}
The above expression is always positive. Therefore,  when $\lambda$ is small ($\ell\approx r_0$), we can conclude that the weak and null energy conditions are satisfied in all regions.

\subsubsection{Strong Condition}

Finally, for the strong condition, we must impose
\begin{equation} \label{strong}
\rho+p_r+2p_t=2p_t\geq 0\to E_s\equiv -\rho-\frac{r}{2}\rho'\geq 0,
\end{equation}
where we have used \ref{lateral}. The expression for $E_s$ becomes 
\begin{align}
    E_s=&-\frac{1}{4 r_c^3}\text{sech}^2\left(\frac{\pi  r^3 }{2 r_c^3}\sqrt{1-\frac{r_c^6}{r^6}}\right) \left(3 \pi  \lambda  r^3 \sinh \left(\frac{\pi  \lambda  r^3}{r_c^3}\right)+4 r_c^3\right)\nonumber\\
    &+\frac{3 \pi  r^3 }{2 r_c^3 \sqrt{1-\frac{r_c^6}{r^6}}}\tanh \left(\frac{\pi  r^3 }{2 r_c^3}\sqrt{1-\frac{r_c^6}{r^6}}\right) \text{sech}^2\left(\frac{\pi  r^3 }{2 r_c^3}\sqrt{1-\frac{r_c^6}{r^6}}\right) \sinh ^2\left(\frac{\pi  \lambda  r^3}{2 r_c^3}\right)\label{fullstrong}   
\end{align}
With the above expression, we can analyze the behavior of $E_s$ in different  regions. 

In the regime $r\gg r_c$ (small $\alpha$), we get from (\ref{fullstrong}) 
\begin{equation}
E_s=\rho\left(-1+\frac{3 }{2}\frac{r^3}{r_c^3}+\frac{3 }{2}\frac{\alpha a}{r^3}\right)\geq0.
\end{equation}
If we multiply by $2a^3r^3/3$ and define $u=r^3$, the above equation reduces to
\begin{equation}
h(u)\equiv u^2-\frac{2a^3}{3}u+\alpha a^4\geq 0
\end{equation}
In the pure Dymnikova case, we have $\alpha=0$. We get that $h(u)$ is negative for $u=r^3<2a^3/3$. Therefore, the strong energy condition will always be violated as we approach $r=0$. Let us analyze if the  situation with $\alpha\neq 0$ is different.  The roots of the above equation are 
\begin{equation}
u_{\pm}=\frac{2a^3}{6}\left(1\pm\sqrt{1-9\frac{\alpha}{a^2}}\right).
\end{equation}
Since $\alpha$ is very small we will always have that  the above roots are real and the strong energy condition is violated for $u<u_+$. 

Let us analyze the case close to $r_c$. For this, we expand the above expression in $\xi=r-r_c$ to get
\begin{equation} \label{linearweak}
   E_w(r)=c_3+c_4\xi 
\end{equation}
with
\begin{equation}\label{astrong}
c_3=    -\frac{3}{4}  \pi  \lambda\sinh (\pi  \lambda )+\frac{1}{4} \left(3 \pi ^2-4\right) \sinh ^2\left(\frac{\pi  \lambda }{2}\right).
\end{equation}
and 
\begin{align}\label{bstrong}
&c_4=-\frac{3 \pi ^2 }{4 r_c}(\pi ^2-4)-\frac{3\pi^2 }{4 r_c} \left(3 \lambda ^2+\pi ^2-4\right) \cosh (\pi  \lambda )\nonumber \\
&+\frac{ 3\pi\lambda  }{4 r_c}\left(3 \pi ^2-5\right)\sinh (\pi  \lambda )
\end{align}
From (\ref{astrong}) and remembering that $0\leq\lambda\leq 1$, we always have that $c_3>0$. Therefore, the strong energy condition is satisfied for $\xi=0$ ($r=r_c$). 

Let us now analyze if there is some region in which the conditions are violated. Since (\ref{linearweak}) is linear and $c_3>0$, the conditions are violated for $\xi>-c_3/c_4$, if $c_4<0$. We must be careful when analyzing $c_4$ as a function of $\lambda$. We should remember that both, $\lambda$ and $r_c$, depend on $\alpha$. Restoring this in expression \ref{bstrong} we have that, close to $\lambda=1$ we have 
$$
c_4\propto -\frac{1}{\alpha^{1/3}} \to \lim_{\alpha \to 0}c_4=-\infty
$$ 
Therefore, in  the limit $\alpha \to 0$, the conditions are violated for all $r$. This is precisely the pure Dymnikova limit that we advanced above.  However, when $\lambda=0$ (and therefore the upper value of $\alpha$), we get $c_4\to 0 $, and the conditions are never violated in the regimen $r_c=r_0$. Therefore the limit $\lambda=0$ is very important, and we must analyze if this conclusion is valid only close to $r=r_c$. For this, we must use (\ref{fulldensityupper}) in (\ref{strong}) to get 
\begin{equation}
E_s=\frac{\lambda ^2 \text{sech}^2\left(\frac{\pi  r^3 }{2 r_c^3}\sqrt{1-\frac{r_c^6}{r^6}}\right)}{8 r_c^9 \sqrt{1-\frac{r_c^6}{r^6}}} \left(3 \pi ^3 r^9 \tanh \left(\frac{\pi  r^3 }{2 r_c^3}\sqrt{1-\frac{r_c^6}{r^6}}\right)-8 \pi ^2 r_c^3 r^6 \sqrt{1-\frac{r_c^6}{r^6}}\right)    
\end{equation}
The above expression is always positive. Therefore,  when $\lambda$ is small ($\ell\approx r_0$), we see that the strong energy condition is satisfied everywhere. Thus, our GUP-corrected solution represents a regular configuration in which the strong energy condition is satisfied in all regions. 

\begin{figure}[!h]
 \centering
     \includegraphics[width=0.9\textwidth]{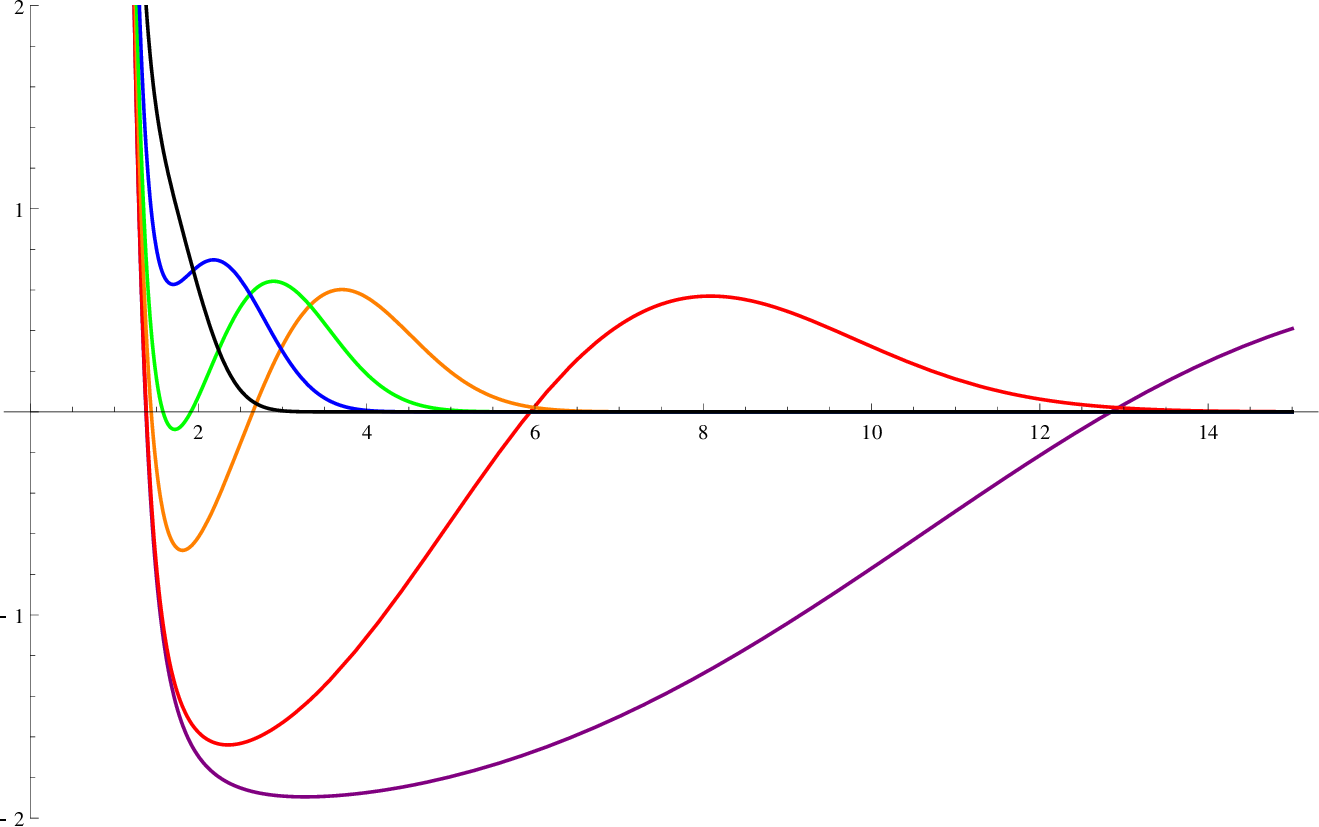}
       \caption{Representation of the strong energy condition associated to the Dyminikova-GUP black hole, for $\lambda=0.9999$ (purple),  $\lambda=0.999$ (red),  $\lambda=0.99$ (orange),  $\lambda=0.98$ (green),  $\lambda=0.96$ (blue), and  $\lambda=0.90$ (black). The strong energy condition is always satisfied around $r_c$. When $\lambda \to 1$, the condition is violated in the de Sitter core region, but it is satisfied for configurations in which the GUP scale and the de Sitter radius are sufficiently close ($\lambda$ not too close to unity). The green line ($\lambda=0.98$) is about to satisfy the condition. The bigger the size of $r_0$ as compared to the GUP scale, the bigger the region where this condition is violated. In the $x$ axis we have used $r_c=1$. }
   \label{strongfig}
   \end{figure}

Finally, in Fig. \ref{strongfig} we plot the strong energy condition for various configurations using the exact expression for the energy density and pressures. 
The plot illustrates the sum of density and pressures associated with the Dyminikova-GUP black hole for several values of $\lambda$. An important observation from this figure  is that the strong condition is never violated around the minimal radius $r_c$ ($x=1$). However, as we move farther away from the origin, this condition can indeed be violated in the Dymnikova limit, when $\lambda\to 1$. 
This region diminishes as the relative size between the GUP and core scales become comparable. It is worth noting that the region of non-violation of this condition diminishes as the minimum GUP length decreases. In other words, smaller values of $\alpha$ correspond to a smaller extent of the region where the strong conditions are not violated. Thus, as $\alpha$ vanishes, the violation of the strong condition extends across the entire spacetime, reinforcing the contrast between the Dyminikova-GUP black hole and the regular Dymnikova black hole.

\section{Conclusion}

In this work we have investigated the impact of quantum corrections on the Dymnikova regular black hole induced by incorporating a new length scale inspired by the Generalized Uncertainty Principle (GUP). By introducing that minimum length, the GUP modifies the energy density associated with the gravitational source, known as the Dymnikova vacuum, based on the gravitational analogue of the Schwinger effect. Our analysis provides novel solutions encompassing a broad spectrum of black hole sizes and regular naked objects, ranging from microscopic to macroscopic scales. 

We have derived Dymnikova-GUP black hole solutions by considering small minimum length approximations at both large and small radial distances. Our analysis revealed the presence of two horizons with the inner horizon larger than in the absence of GUP, while the outer horizon practically coincides with Dymnikova's solution (and with Schwarzschild's one), as our analytical approximations explicitly show. Importantly, these GUP-modified solutions remain regular and devoid of singularities at all scales, mirroring the regularity of the original Dymnikova solution. A key difference with the Dymnikova case is the generic emergence of a wormhole in the innermost region of the object. For small configurations, this wormhole can be traversable (without horizons), though in general, it is hidden behind two horizons, like in electrically charged solutions. 

Furthermore, an analytical exploration of the energy conditions across all scales revealed an intriguing observation. While the pure Dymnikova case is characterized by a violation of the strong, weak, and null energy conditions, we have found that these violations do not occur at the shortest scales for the GUP corrected solution. This unexpected contrast suggests a departure from conventional expectations and highlights the significant influence of quantum corrections and the GUP in modifying the energy conditions near the Planck scale.

Overall, this study contributes to our understanding of the behavior of regular black holes in the presence of quantum corrections and the GUP. Our findings support the notion that incorporating quantum effects can lead to solutions that are free from singularities and exhibit novel features regarding energy conditions and topology. These results pave the way for further investigations into the interplay between quantum mechanics and gravity, providing valuable insights into the nature of black holes and the fundamental structure of spacetime.

\acknowledgments The authors G. Alencar and C. R. Muniz would like to thank Conselho Nacional de Desenvolvimento Cient\'{i}fico e Tecnol\'{o}gico (CNPq) and Fundação Cearense de Apoio ao Desenvolvimento Científico e
Tecnológico (FUNCAP) through PRONEM PNE0112- 00085.01.00/16, for the partial financial support. Milko Estrada is funded by the FONDECYT Iniciaci\'on Grant 11230247. This work is also supported by the Spanish Grant PID2020-116567GB- C21 funded by MCIN/AEI/10.13039/501100011033, and the project PROMETEO/2020/079 (Generalitat Valenciana). Further support is provided by the EU's Horizon 2020 research and innovation (RISE) programme H2020-MSCA-RISE-2017 (FunFiCO-777740) and  by  the  European Horizon  Europe  staff  exchange  (SE)  programme HORIZON-MSCA-2021-SE-01 (NewFunFiCO-10108625).


\begin{thebibliography}{99}

\bibitem{Schwar}  
K.~Schwarzschild, Sitzungsberichte der Königlich Preussischen Akademie der Wissenschaften. {\bf 7}, 189 (1916).

\bibitem{Oppy}
J.~R.~Oppenheimer and H.~Snyder
Phys. Rev. {\bf56}, 455 (1939).

\bibitem{Korm}
J.~Kormendy and K.~Gebhardt, AIP Conference Proceedings, {\bf586} (1): 363 (2001). doi: 10.1063/1.1419581.

\bibitem{Ragavendra:2023ret}
H.~V.~Ragavendra, H. V. and L.~Sriramkumar,
 Galaxies, {\bf 11}, 1, 34, (2023),
 doi: 10.3390/galaxies11010034,
arXiv: 2301.08887 [astro-ph.CO].

\bibitem{Kazun}
A.~Kazunori Akiyama {\it et al.},
Astrophys.J.Lett. 875 (2019) L1, DOI: 10.3847/2041-8213/ab0ec7; Astrophys.J.Lett. 875 (2019) 1, L2, DOI: 10.3847/2041-8213/ab0c96; Astrophys.J.Lett. 875 (2019) 1, L3, DOI: 10.3847/2041-8213/ab0c57; Astrophys.J.Lett. 875 (2019) 1, L4, DOI: 10.3847/2041-8213/ab0e85; Astrophys.J.Lett. 875 (2019) 1, L5, DOI: 10.3847/2041-8213/ab0f43; Astrophys.J.Lett. 875 (2019) 1, L6, DOI: 10.3847/2041-8213/ab1141.

\bibitem{Kazun1}
A.~Kazunori {\it et al.}, Astrophys.J.Lett. 930 (2022) 2, L12, DOI: 10.3847/2041-8213/ac6674; Astrophys.J.Lett. 930 (2022) 2, L13, DOI: 10.3847/2041-8213/ac6675; Astrophys.J.Lett. 930 (2022) 2, L14, DOI: 10.3847/2041-8213/ac6429; Astrophys.J.Lett. 930 (2022) 2, L15, DOI: 10.3847/2041-8213/ac6736.

\bibitem{Vagnozzi:2022moj}
S.~Vagnozzi {\it et al.}, arxiv: 2205.07787 [gr-qc], rep.Numb. UCI-HEP-TR-2022-07, 5 (2022).

\bibitem{Steer}
S.~Mastrogiovanni, D.~A.~Steer, and M.~Barsuglia,
Phys. Rev. {\bf D102}, 044009 (2020),
doi: 10.1103/PhysRevD.102.044009.

\bibitem{Elias}
E.~C.~Vagenas, S.~M.~Alsaleh, and A.~F.~Ali,
EPL, {\bf120}, 40001 (2017),
doi: 10.1209/0295-5075/120/40001.

\bibitem{Amati1}
D.~Amati, M.~Ciafaloni, and G.~Veneziano G., Phys.
Lett. {\bf B197}, 81 (1987).

\bibitem{Gross}
D.~J.~Gross and P.~F.~Mende,
Phys. Lett. {\bf B197}, 129 (1987).

\bibitem{Amati2}
D.~Amati, M.~Ciafaloni, G.~Veneziano, Phys. Lett. {\bf B216}, 41 (1989).


\bibitem{Paffuti}
K.~Konishi, G.~Paffuti, and P.~Provero P., Phys. Lett.
{\bf B234}, 276 (1990).

\bibitem{Nozari}
P.~Pedram, K.~Nozari, and S.~H.~Taheri, JHEP {\bf03}, 093 (2011).

\bibitem{Bardeen}
J.~M.~Bardeen, Conference Proceedings of GR5, Tbilisi,
URSS, 174 (1968).

\bibitem{Frolov:2016pav}
V.~P.~Frolov,
   Phys. Rev. {\bf D94}, 10, 104056 (2016).
   doi: 10.1103/PhysRevD.94.104056.

  \bibitem{Ali}
   A.~H.~Chamseddine, V.~Mukhanov, Eur.Phys.J. {\bf C77}, 3, 183 (2017),
   doi: 10.1140/epjc/s10052-017-4759-z.

   \bibitem{Dymnikova:1992ux}
I.~Dymnikova,
Gen. Rel. Grav. \textbf{24}, 235-242 (1992)
doi:10.1007/BF00760226.


\bibitem{Olmo:2015dba}
G.~J.~Olmo, D.~Rubiera-Garcia and A.~Sanchez-Puente,
Eur. Phys. J. C \textbf{76}, no.3, 143 (2016)
doi:10.1140/epjc/s10052-016-3999-7
[arXiv:1504.07015 [hep-th]].

\bibitem{Olmo:2015axa}
G.~J.~Olmo and D.~Rubiera-Garcia,
Universe \textbf{1}, no.2, 173-185 (2015)
doi:10.3390/universe1020173
[arXiv:1509.02430 [hep-th]].

\bibitem{Menchon:2017qed}
C.~Menchon, G.~J.~Olmo and D.~Rubiera-Garcia,
Phys. Rev. D \textbf{96}, no.10, 104028 (2017)
doi:10.1103/PhysRevD.96.104028
[arXiv:1709.09592 [gr-qc]].


\bibitem{Bejarano:2017fgz}
C.~Bejarano, G.~J.~Olmo and D.~Rubiera-Garcia,
Phys. Rev. D \textbf{95}, no.6, 064043 (2017)
doi:10.1103/PhysRevD.95.064043
[arXiv:1702.01292 [hep-th]].


\bibitem{Eloy}
E.~Ayon-Beato and A.~Garcia,
Phys.Rev.Lett. {\bf80} 5056 (1998),
doi: 10.1103/PhysRevLett.80.5056.

\bibitem{Stephano}
S.~Ansoldi, P~.~Nicolini, A.~Smailagic, and E.~Spallucci,
Phys.Lett. {\bf B645}, 261 (2007),
doi: 10.1016/j.physletb.2006.12.020.

\bibitem{Bazeia:2015uia}
D.~Bazeia, L.~Losano, G.~J.~Olmo, D.~Rubiera-Garcia and A.~Sanchez-Puente,
Phys. Rev. D \textbf{92}, no.4, 044018 (2015)
doi:10.1103/PhysRevD.92.044018
[arXiv:1507.07763 [hep-th]].

\bibitem{Maluf:2018ksj}
R.~V.~Maluf and J.~C.~S.~Neves,
   Int. J. Mod. Phys. {\bf D28}, 03, 1950048 (2018),  
    doi: 10.1142/S0218271819500482.

   \bibitem{Maluf:2022jjc}
R.~V.~Maluf, C.~R.~Muniz, A.~C.~L.~Santos, and M.~Estrada,
   Phys. Lett. {\bf B835}, 137581 (2022),
   doi: 10.1016/j.physletb.2022.137581.

   \bibitem{Maluf:2022qfc}
 R.~V.~Maluf, C.~R.~Muniz, A.~C.~L.~Santos,
  Astrophys. Space Sci. {\bf367}, 9, 90 (2022),  
    doi:10.1007/s10509-022-04118-6.

\bibitem{Hugo}
H.~Christianssen {\it et al.},
Accepted in intl. J. Mod. Phys. D,
doi: 10.1142/S0218271823500414.

\bibitem{Simpson:2018tsi}
A.~Simpson and M.~Visser,
JCAP \textbf{02}, 042 (2019)
doi:10.1088/1475-7516/2019/02/042
[arXiv:1812.07114 [gr-qc]].

%
\bibitem{BeltranJimenez:2017doy}
J.~Beltran Jimenez, L.~Heisenberg, G.~J.~Olmo and D.~Rubiera-Garcia,
Phys. Rept. \textbf{727} (2018), 1-129
doi:10.1016/j.physrep.2017.11.001
[arXiv:1704.03351 [gr-qc]].





\bibitem{Furtado:2022tnb}
J.~Furtado and G.~Alencar,
Universe \textbf{8}, no.12, 625 (2022)
doi:10.3390/universe8120625
[arXiv:2210.06608 [gr-qc]].

\bibitem{Lima:2022pvc}
A.~M.~Lima, G.~M.~de Alencar Filho and J.~S.~Furtado Neto,
Symmetry \textbf{15}, no.1, 150 (2023)
doi:10.3390/sym15010150
[arXiv:2211.12349 [gr-qc]].

\bibitem{Gliner1966}
Gliner, E. B., Soviet Physics JETP, \textbf{22}, 378 (1966)

\bibitem{Gliner1970}
Gliner, E. B., Soviet Physics Doklady, \textbf{15}, 559 (1970)



    
\bibitem{Dymnikova:1996plb}
I.~G.~Dymnikova,
Int.J.Mod.Phys. {\bf D5}, 5, 529 (1996),
doi: 10.1142/s0218271896000333.



\bibitem{Ansoldi:2008jw}
S.~Ansoldi,
Conference on Black Holes and Naked Singularities,
reportNumber: KUNS-2108, 2 (2008),
[arXiv: 0802.0330 [gr-qc]].


\bibitem{Haouat:2013yba}
S.~Haouat and K.~Nouicer,
Phys. Rev. {\bf D89}, 10, 105030 (2014),
 doi: 10.1103/PhysRevD.89.105030,
[arXiv: 1310.6966 [hep-th]].


\bibitem{Ong:2020tvo}
Y.~C.~Ong,
Eur. Phys. J. {\bf C80}, 8, 777 (2020),
    doi: 10.1140/epjc/s10052-020-8363-2,
    [arXiv: 2005.12075 [gr-qc]].

\bibitem{Estrada:2023pny}
M.~Estrada and C.~R.~Muniz,
JCAP {\bf03}, 055 (2023),  
doi = 10.1088/1475-7516/2023/03/055.    


\bibitem{Kastor:2008xb}
D.~Kastor,
Class. Quant. Grav. \textbf{25}, 175007 (2008)
doi:10.1088/0264-9381/25/17/175007

\bibitem{Aros:1999kt}
R.~Aros, M.~Contreras, R.~Olea, R.~Troncoso and J.~Zanelli,
Phys. Rev. D \textbf{62}, 044002 (2000)
doi:10.1103/PhysRevD.62.044002
[arXiv:hep-th/9912045 [hep-th]].

\bibitem{Wald:1993nt}
R.~M.~Wald,
Phys. Rev. D \textbf{48}, no.8, R3427-R3431 (1993)
doi:10.1103/PhysRevD.48.R3427
[arXiv:gr-qc/9307038 [gr-qc]]

\bibitem{Zhang:2020mxi}
H.~X.~Zhang, Y.~Chen, T.~C.~Ma, P.~Z.~He and J.~B.~Deng,
Chin. Phys. C \textbf{45}, no.5, 055103 (2021)
doi:10.1088/1674-1137/abe84c

\bibitem{Aros:2019quj}
R.~Aros and M.~Estrada,
Eur. Phys. J. C \textbf{79}, no.3, 259 (2019)
doi:10.1140/epjc/s10052-019-6783-7
[arXiv:1901.08724 [gr-qc]].

\bibitem{Estrada:2020tbz}
M.~Estrada and F.~Tello-Ortiz,
EPL \textbf{135}, no.2, 20001 (2021)
doi:10.1209/0295-5075/ac0ed0
[arXiv:2012.05068 [gr-qc]].

\bibitem{Ma:2014qma}
M.~S.~Ma and R.~Zhao,
Class. Quant. Grav. \textbf{31}, 245014 (2014)
doi:10.1088/0264-9381/31/24/245014
[arXiv:1411.0833 [gr-qc]].

\bibitem{Penrose:1964wq}
R.~Penrose,
Phys. Rev. Lett. \textbf{14}, 57-59 (1965)
doi:10.1103/PhysRevLett.14.57
\end{thebibliography}
\end{document}